\documentclass[aps,prl,reprint,groupedaddress]{revtex4-1}

\usepackage{graphicx}% Include figure files
\graphicspath{figures/}
\usepackage{dcolumn}% Align table columns on decimal point
\usepackage{amssymb}
\usepackage{amsmath}
\usepackage{fancyhdr}
\usepackage{bm}% bold math
\usepackage[colorlinks,linkcolor=blue,anchorcolor=blue,citecolor=blue,urlcolor=blue]{hyperref}

\begin{document}
\draft
\title{Quantum-enhanced sensing of photonic modes with cat states}
\author{Xiao-Wei Zheng$^{1}$}
\author{Jun-Cong Zheng$^{1}$}
\author{Xue-Feng Pan$^{1}$}
\author{Pengbo Li$^{1}$$^\ast$}
\address{$^{1}$School of Physics, Xi'an Jiaotong University, Xi'an 710049,\\
	China\\ }
\date{\today }

\begin{abstract}
Quantum coherence is critical resource for applications in quantum
technology, among which quantum-enhanced sensing represents a typical
example. Compared with quantum metrology with entangled states of multiple
qubits, bosonic interferometers have the advantage of being
hardware-efficient, enabled by exploiting the high-dimensional Hilbert space
of a bosonic mode. Phase estimations with precisions approaching the
Heisenberg limit have been demonstrated with superpositions of highly
nonclassical Fock states of a single bosonic mode. We here present a scheme
for realizing quantum metrology based on a fundamentally distinct kind of
superposition states$-$Schr\"{o}dinger cat states, defined as quantum
superpositions of two quasiclassial (coherent) states of a bosonic mode. The
interferometer aims to estimate the frequency shift of the photonic mode with
respect to a reference frequency. The phase-space rotation due to such a
frequency shift, together with a displacement operation, produces a phase
difference between these two superposed quasiclassical state components,
that can be extracted with a qubit. The signal-to-noise ratio exhibits a
scaling approaching the Heisenberg limit, like interferometers based on Fock
state superpositions, but without requiring a step-by-step procedure to
prepare the resource state.
\end{abstract}

\vskip0.5cm

\narrowtext
\maketitle
\section{INTRODUCTION}

\bigskip The capability of building a high-sensitive interferometer for
performing a high-precision measurement on physical quantity is crucial for
advancement of science and technology [1-5]. In a standard interferometer,
the physical quantity to be measured is encoded in a phase shift $\theta $
in one arm of the interferometer, and then estimated by interferometric
techniques. The uncertainty of the phase estimation $\Delta \theta $ can
arise from classical noises or quantum fluctuations. One strategy to
mitigate the influence of noises is to prepare N identical interfering
particles and average over all the measurement outcomes, with the precision
scaling with N in a manner depending whether and how the particles are
correlated. When these particles are not correlated or classical correlated,
the phase uncertainty is reduced by a factor proportional to $1/\sqrt{N}$,
as exemplified by a coherent state of a light field with N photons that
exhibits no quantum effects. Such a standard quantum limit (SQL) can be
surpassed when these N particles are prepared in nonclassically correlated
states, among which the $N$-qubit Greenberger--Horne--Zeilinger (GHZ) state
[6] is a paradigm, with which the precision exhibits a $1/N$ scaling, known
as Heisenberg limit (HL) [5].

So far, entanglement-enhanced metrology has been demonstrated in a variety
of platforms, including optical systems [7-12], Bose-Einstein Condensates
[13-17], trapped ions [18-22], nuclear magnetic resonance [23], a
superconducting circuit [24], and an atom array trapped in an optical
lattice [25]. In principle, the sensitivity can be improved with the
increase of the size of the entangled state, but which requires more
hardware overhead, including the entangled elements and corresponding
readout apparatuses. Fortunately, entanglement is not the necessary resource
for quantum-enhanced metrology [5,26]. Recently, superpositions of Rydberg
states of single atoms were demonstrated to be useful for realizing
high-sensitivity probes beyond SQL [27-29]. Bosonic modes represent an
alternative promising candidate for implementation of quantum-enhanced
metrology in a resource-efficient manner, enabled by their infinite
dimensional Hilbert space. Single-mode interferometers have been built with
a mechanical oscillator [30] and with a photonic field stored in a microwave
cavity [31], where HL was approached by exploiting the quantum coherence
between the ground state ($\left\vert 0\right\rangle $) and a
high-quantum-number Fock state ($\left\vert N\right\rangle $) of the bosonic
mode. Quantum-enhanced metrology with Fock states has also been demonstrated
[32,33]. Generally, preparation of such nonclassical probe states with a
large N requires a step-by-step procedure [30] or a projective measurement,
which converts a quasiclassical coherent state into the desired quantum
state in a probabilistic manner [33,34].

Cat states of bosonic modes represent another kind of typical nonclassical
states [35]. Such states are formed by coherent states with different
amplitudes or phases. Although coherent states themselves are robust
quasiclassical components, cat states can exhibit strong nonclassical
features, as a consequence of the quantum interference effects between the
quasiclassical components. The tunnelling probability between two
superimposed quasiclassical components is exponentially suppressed with the
increase of their distance in phase space [36,37]. This robustness makes cat
states a promising candidate for construction of inherently-protected qubits
for encoding quantum information [38-40]. Metrological power of such states
has been theoretically analyzed from the viewpoint of quantum resources
[41], and its application in the high-precision detection of a phase-space
displacement has been investigated in both theory [42,43] and experiment
[44,45].  The authors of Ref. [43] also proposed a protocol for
quantum-enhanced measurement of rotations of a bosonic mode, where an
entangled mesoscopic state involving the bosonic mode and a qubit serves as
the quantum resource. After the rotation, the qubit-field entanglement is
undone by their interaction, but the field evolution trajectories associated
with the two qubit states accumulate a phase difference, which can be
measured by measuring the qubit's population. The sensitivity scales with
the phase-space separation between the two photonic coherent states of the
initial mesoscopic entanglement. As the field's rotation is produced by the
deviation of the frequency of the photonic field from a reference frequency,
the field frequency can be inferred from the rotation angle. The protocol is
valid only when the phase uncertainty of the qubit is negligible during the
field's rotation. However, in real physical systems (e.g., circuit QED
architectures [46]), the dephasing time of the qubit may be much shorter
than that of the photonic mode. As a consequence, the quantum advantage may
be deteriorated by the phase noise of the qubit accumulated during the
field's rotation.

We here present an alternative protocol for estimating the frequency shift
of a photonic mode with the superposition of an amplitude cat state, formed
by the vacuum state and a coherent state. Such a frequency shift can result
in a rotation of the coherent state in phase space. This signal can be
encoded to the state of a qubit by performing a phase-space displacement on
the photonic mode, and then sandwiching a qubit-state-dependent $\pi $-phase
shift of the photonic mode between two $\pi /2$ pulses applied to the qubit.
In distinct contrast with the protocol of Ref. [43], the photonic mode is
not entangled with the qubit during the rotation, so that the purity of the
superposition state is not deteriorated by the qubit dephasing during this
stage. The numerical results show that the precision of phase estimation is
improved with the increase of the cat size, with the scaling approaching the
HL. We further show that this protocol can be adapted to the estimation of
the frequency of mechanical oscillators. The most remarkable feature of this
metrological method is that the preparation of a cat state is much more easy
than that of the Fock state or its superposition with the vacuum state with
the same size. It can be created deterministically with a procedure where
the number of steps does not increase with the cat size [46].

\section{THE PROTOCOL}

\bigskip Let us begin by illustrating how a coherent state $\left\vert
\alpha _{0}\right\rangle $ can be used to amplifying the rotation produced
by a frequency shift of a bosonic mode. In the framework rotating at a
reference frequency $\omega _{0}$, the system evolution is governed by the
free Hamiltonian 
\begin{equation}
H=\varepsilon a^{\dagger }a,
\end{equation}%
where $\varepsilon $ represents the frequency shift with respect to a
reference frequency, and $a^{\dagger }$ and $a$ denote the creation and
annihilation operators for the bosonic mode. This frequency shift results in
a phase-space rotation, evolving the coherent state $\left\vert \alpha
_{0}\right\rangle $ to $\left\vert \alpha _{0}e^{i\theta }\right\rangle $,
where $\theta =-\varepsilon T$ with $T$ being the evolution time. The
quantum-mechanical phase difference between this rotated coherent state and
the original one is defined as the argument of the inner product is $\phi
=\arg \left\langle \alpha _{0}\right\vert \left. \alpha _{0}e^{i\theta
}\right\rangle =D\sin \theta $, where $D=\left\vert \alpha _{0}\right\vert
^{2}$ is mean quantum number of the coherent state $\left\vert \alpha
_{0}\right\rangle $. When $\theta \ll 1$, $\phi \simeq D\theta $. This
implies that the quantum-mechanical phase accumulated during the time $T$ is
proportional to $D$.

\bigskip To detect this frequency-shift-induced phase, it is necessary to
prepare the bosonic mode in the cat state%
\begin{equation}
{\cal N}(\left\vert 0\right\rangle +\left\vert \alpha _{0}\right\rangle ),
\end{equation}%
where $\left\vert 0\right\rangle $ denotes the vacuum component which serves
as a reference state and does not evolve under the phase-space rotation, and 
${\cal N}=(1+e^{-D/2})^{-1/2}/\sqrt{2}$ is the normalization factor. As the
two superimposed quasiclassical state components differ from each other by
the amplitude, the superposition state is referred to as an amplitude cat
state. Under the application of the Hamiltonian of Eq. (1) for a time $T$,
this state evolves to ${\cal N}(\left\vert 0\right\rangle +\left\vert \alpha
_{0}e^{i\theta }\right\rangle )$. Since any coherent state has a zero phase
difference relative to the vacuum component, a phase-space displacement
operation $D(-\alpha _{0}/2)$ is needed to produce an observable phase
difference. This displacement respectively transforms $\left\vert
0\right\rangle $ and $\left\vert \alpha _{0}e^{i\theta }\right\rangle $ to $%
\left\vert -\alpha _{0}/2\right\rangle $ and $e^{iD\theta /2}\left\vert
\alpha _{0}^{\prime }/2\right\rangle $, where $\alpha _{0}^{\prime }=\alpha
_{0}(2e^{i\theta }-1)$. Consequently, the cat state is evolved to 
\begin{equation}
{\cal N}(\left\vert -\alpha _{0}/2\right\rangle +e^{iD\theta /2}\left\vert
\alpha _{0}^{\prime }/2\right\rangle ).
\end{equation}%
When $\theta \ll 1$, the two coherent states approximately have the same
amplitude, and differ from each other by their phases. The resulting
superposition state is referred to as the phase cat state. The corresponding
phase difference can be detected by the Ramsey interferometry, realized by
sandwiching a conditional $\pi $-phase shift $G_{\pi }=(-1)^{a^{\dagger
}a}\left\vert e\right\rangle \left\langle e\right\vert $ between two $\pi /2$
qubit rotation, $e^{-i\pi \sigma _{y}/4}$, where $\sigma _{y}=i\left\vert
e\right\rangle \left\langle g\right\vert -i\left\vert g\right\rangle
\left\langle e\right\vert $ with $\left\vert e\right\rangle $ and $%
\left\vert g\right\rangle $ denoting the qubit's upper and lower levels,
respectively. The conditional $\pi $-phase shift $G_{\pi }$ is realized by
dispersively coupling the bosonic mode to the qubit [46]. The dispersive
interaction is described by the Hamiltonian $H_{I}=-\chi a^{\dagger
}a\left\vert e\right\rangle \left\langle e\right\vert $, where $\chi $
denotes the dispersive coupling strength between the qubit and the bosonic
mode. With the choice of the interaction time $\pi /\chi $, the dispersive
interaction results in $\pi $-phase shift to the qubit's state $\left\vert
e\right\rangle $ conditional on the parity of the quantum number of the
bosonic mode being odd.

The qubit and the bosonic mode is evolved to an entangled state at the
output of the Ramsey interferometer, given by%
\begin{eqnarray}
&&\frac{{\cal N}}{\sqrt{2}}[(\left\vert -\alpha _{0}/2\right\rangle
+e^{iD\theta /2}\left\vert \alpha _{0}^{\prime }/2\right\rangle )\left\vert
\psi _{+}\right\rangle   \nonumber \\
&&+(\left\vert \alpha _{0}/2\right\rangle +e^{iD\theta /2}\left\vert -\alpha
_{0}^{\prime }/2\right\rangle )\left\vert \psi _{-}\right\rangle ,
\end{eqnarray}%
where%
\begin{equation}
\left\vert \psi _{\pm }\right\rangle =\left( \left\vert e\right\rangle
-\left\vert g\right\rangle \right) /\sqrt{2}.
\end{equation}%
\begin{figure*}[htbp]
	\includegraphics[width=7.1in]{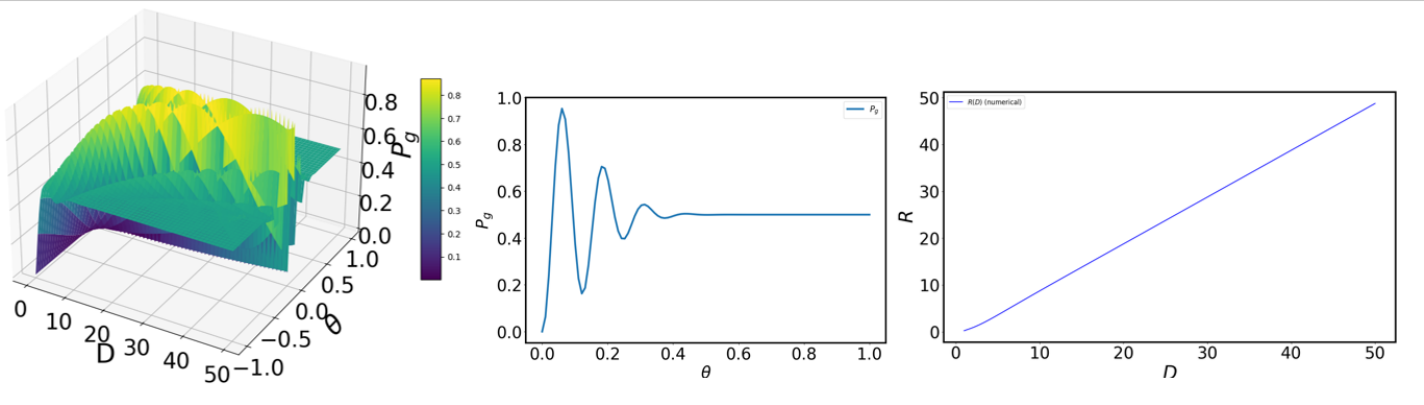} 
	\caption{ Quantum sensing without decoherence. (a) The qubit's $%
		\left\vert g\right\rangle $-state population ($P_{g}$) as a function of $D$
		and $\theta $. (b) $P_{g}$ as a function of $\theta $ for $D=50$. \bigskip
		(c) Signal-to-noise ratio ($R$) as a function of $D$ at the bias point $%
		\theta _{0}=\pi /(2D)$.}
\end{figure*}
This result shows that the Ramsey interferometer splits each of the two
input quasiclassical component into two components with opposite phases. The
resulting quantum interference leads to the $\theta $-dependence of the
qubit's $\left\vert g\right\rangle $-state population, 
\begin{eqnarray}
P_{g} &=&{\cal N}^{2}\{1-(e^{-D/2}+e^{-D^{\prime }/2})/2  \nonumber \\
&&+e^{-(D+D^{\prime })/8}%
%TCIMACRO{\func{Re}}%
%BeginExpansion
\mathop{\rm Re}%
%EndExpansion
[e^{iD\theta /2}(e^{-\alpha _{0}^{\ast }\alpha _{0}^{\prime }/4}-e^{\alpha
_{0}^{\ast }\alpha _{0}^{\prime }/4})]\}
\end{eqnarray}%
where $D^{\prime }=\left\vert \alpha _{0}^{\prime }\right\vert ^{2}$. Fig.
1a shows this probability as a function of $D$ and $\theta $. As expected,
when $D$ is sufficiently large and $\theta $ is sufficiently small, $P_{g}$
approximately exhibits an oscillatory pattern, whose period decreases as $D$
increases. This is due to the fact that $P_{g}$ can be well approximated by $%
{\cal C}(\theta )=[1-\cos (D\theta )]/2$ when $D\gg 1$ and $D\theta
^{2}/2\ll 1$. To show this behavior more clearly, we display $P_{g}$ as a
function of $\theta $ for $D=50$ in Fig. 1b, which confirms the
ultra-sensitive response of $P_{g}$ to slight change of $\theta $ during the
regime $0<\theta <\pi /(2D)$.

The performance of the sensor is characterized by the signal-to-noise ratio,
defined by%
\begin{equation}
R=\frac{\left\vert \partial P_{g}/\partial \theta \right\vert }{\sqrt{%
P_{g}(1-P_{g})}}.
\end{equation}%
The interferometer aims to estimate the deviation of $\theta $ from a bias
point $\theta _{0}$. To make the interferometer be able to distinguish the
sign of deviation, we set $D\theta _{0}=\pi /2$. At such a bias point, the
signal-to-noise ratio can be approximated by $R\simeq D$. This implies that
the interference presents an approximately linearly improved sensitivity
with the increase of the mean photon number, approaching the HL when the
decoherence is negligible. We note that a similar scaling has been
demonstrated with the Fock-state superposition, $(\left\vert 0\right\rangle
+\left\vert N\right\rangle )/\sqrt{2}$ [30,31]. However, the pulse sequence
for preparing such a superposition and for recombining the two Fock-state
components becomes more complex with the increase of $N$. In distinct
contrast, the number of steps for preparing and manipulating the cat state
is generally independent of the size of the cat state [46]. \bigskip To
confirm the validity of the approximation, we perform a numerical simulation
of $R$. \bigskip Fig. 1c presents $R$ as a function of $D$ at the bias point 
$\theta _{0}=\pi /(2D)$. The result shows that the signal-to-noise ratio
almost linearly scales with the size of the cat state.

\section{EFFECT OF DECOHERENCE}

The main problem for the implementation of the our protocol is the photonic
dissipation of the cat state. We suppose that the time needed to prepare the
cat state and that to read out the signal are much shorter than the time
needed to accumulate the phase $\theta $, so that the decoherence during the
cat state preparation and the dispersive coupling can be neglected. With the
decoherence effect during the free evolution of the cat state being taken
into consideration, the system combined by the photonic mode and qubit
finally evolves to the mixed state, described by the density operator

\begin{eqnarray}
\frac{{\cal N}^{2}}{2}[\rho _{p}\otimes \left\vert \psi _{+}\right\rangle
\left\langle \psi _{+}\right\vert  &&+\Pi \rho \Pi \otimes \left\vert \psi
_{-}\right\rangle \left\langle \psi _{-}\right\vert   \nonumber \\
+\Pi \rho \otimes \left\vert \psi _{-}\right\rangle \left\langle \psi
_{+}\right\vert  &&+\rho \Pi \otimes \left\vert \psi _{+}\right\rangle
\left\langle \psi _{-}\right\vert .
\end{eqnarray}%
where $\left\vert \psi _{\pm }\right\rangle $ are given by Eq. (5), $\Pi
=e^{-i\pi a^{\dagger }a}$, and%
\begin{eqnarray}
\rho _{p} &=&{\cal N}^{2}(\left\vert -\alpha _{0}/2\right\rangle
\left\langle -\alpha _{0}/2\right\vert +\left\vert \alpha _{0}^{\prime
\prime }/2\right\rangle \left\langle \alpha _{0}^{\prime \prime
}/2\right\vert   \nonumber \\
&&+(Ke^{-i\phi }\left\vert -\alpha _{0}/2\right\rangle \left\langle \alpha
_{0}^{\prime \prime }/2\right\vert +H.c.),
\end{eqnarray}%
with $K=e^{-\left\vert \alpha _{0}\right\vert ^{2}(1-e^{-\kappa T})/2}$, $%
\phi =\frac{1}{2}\left\vert \alpha _{0}\right\vert ^{2}e^{-\kappa T/2}\sin
\theta $, and $\alpha _{0}^{\prime \prime }=\alpha _{0}(2e^{i\theta -\kappa
T/2}-1)$, with $\kappa $ being the photonic decaying rate. The probability
of detecting the qubit in the state $\left\vert g\right\rangle $ is%
\begin{eqnarray}
P_{g} &=&{\cal N}^{2}\{1-(e^{-D/2}+e^{-D^{\prime \prime }/2})/2  \nonumber \\
&&+Ke^{-(D+D^{\prime \prime })/8}%
%TCIMACRO{\func{Re}}%
%BeginExpansion
\mathop{\rm Re}%
%EndExpansion
[e^{i\phi }(e^{-\alpha _{0}^{\prime \prime }\alpha _{0}^{\ast }/4}-e^{\alpha
_{0}^{\prime \prime }\alpha _{0}^{\ast }/4})]\}
\end{eqnarray}%
where $D^{\prime \prime }=\left\vert \alpha _{0}^{\prime \prime }\right\vert
^{2}$.

Fig. 2a displays this probability as a function of $\theta $ and $\kappa T$
with the choice $D=50$. As expected, when $\kappa T$ and $\theta $ are
sufficiently small, $P_{g}$ also approximately exhibits an oscillatory
pattern, but with a reduced contrast compared to the case without
considering decoherence. The result can be well understood by considering
the conditions $D\gg 1$, $D\theta ^{2}/2\ll 1$, and $\kappa T\ll 1$, under
which $P_{g}$ can be well approximated by $P_{g}=[1-e^{-\left\vert \alpha
_{0}\right\vert ^{2}\kappa T/2}\cos (D\theta )]/2$. Fig. 2b presents the
signal-to-noise ratio $R$ versus $D$ and $\kappa T$ at the bias point $%
\theta _{0}=\pi /(2D)$. Due to the presence of decoherence, $R$ does not
show the linear scaling. There is a tradeoff between the gain from the
enhanced sensitivity of the total phase $D\theta $ and the reduced contrast
of the interference fringe. To illustrate this point more clearly, in Fig.
2c we present $R$ as a function of $D$ with the choice $\kappa T=0.02$. As
expected, $R$ does not monotonously increase with $D$. It reaches the
maximum of 36.25 at the point $D=100$, and then decreases due to the linearly
increasing decoherence rate of the cat state.
\begin{figure*}[htbp]
	\includegraphics[width=7.1in]{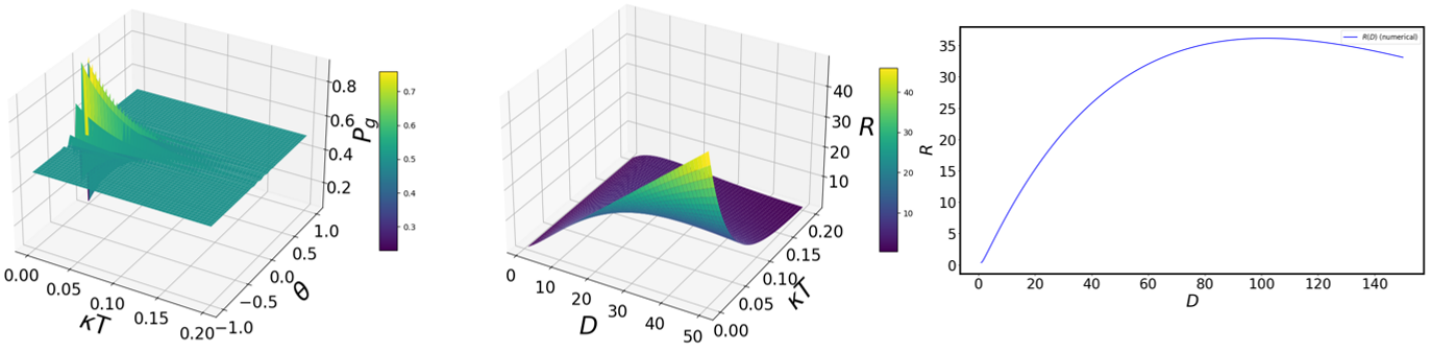} 
	\caption{Quantum sensing with decoherence. (a) The qubit's $%
		\left\vert g\right\rangle $-state population as a function of $\theta $ and $%
		\kappa T$ for $D=50$. (b) Signal-to-noise ratio $R$ versus $D$ and $\kappa T$%
		. (c) $R$ versus $D$ for $\kappa T=0.02$. In both (b) and (c), $R$ is
		calculated at the bias point $\theta _{0}=\pi /(2D)$.}
\end{figure*}
\section{CONCLUSION}

In conclusion, we have proposed a scheme for realizing a quantum-enhanced
interferometer based on an amplitude cat state, formed by the vacuum state
and a coherent state of a photonic mode. The frequency shift of the bosonic
mode with respect to a reference frequency leads to a phase-space rotation
of the coherent state. The rotation angle can be estimated by the Ramsey
interference of a qubit dispersively coupled to the bosonic mode, following
a displacement operation, transforming the amplitude cat state into a phase
cat state. The Ramsey interferometry invovles two $\pi /2$ pulses, in
between which a qubit-state-dependent $\pi $-phase shift of the bosonic mode
is sandwiched. The output qubit's ground state population depends upon the
rotation angle of the bosonic mode, with the sensitivity of being
proportional to the size of the cat state. Our results opens a promising
prospect for quantum metrology in a resource-efficient manner.

This work was supported by the National Natural Science Foundation of China
under Grant No..

Fig. 1 (color online). Quantum sensing without decoherence. (a) The qubit's $%
\left\vert g\right\rangle $-state population ($P_{g}$) as a function of $D$
and $\theta $. (b) $P_{g}$ as a function of $\theta $ for $D=50$. \bigskip
(c) Signal-to-noise ratio ($R$) as a function of $D$ at the bias point $%
\theta _{0}=\pi /(2D)$.

Fig. 2 (color online). Quantum sensing with decoherence. (a) The qubit's $%
\left\vert g\right\rangle $-state population as a function of $\theta $ and $%
\kappa T$ for $D=50$. (b) Signal-to-noise ratio $R$ versus $D$ and $\kappa T$%
. (c) $R$ versus $D$ for $\kappa T=0.02$. In both (b) and (c), $R$ is
calculated at the bias point $\theta _{0}=\pi /(2D)$.

\end{document}